\documentstyle[12pt]{article}
\newcommand{\bea}{\begin{eqnarray}}
\newcommand{\eea}{\end{eqnarray}}

\newcommand{\ee}{\end{equation}}
\newcommand{\be}{\begin{equation}}

\begin{document}

\title{The Goldstone model static solutions on $S^1$}

\author{{\large Y. Brihaye}$^{\diamond}$
and {\large T.N. Tomaras}$^{\dagger}$
\\
$^{\diamond}${\small Department of Mathematical Physics, University 
of Mons-Hainaut, Mons, Belgium}\\
$^{\dagger}${\small 
Department of Physics and Institute of Plasma Physics, University of Crete,} \\
{\small and FO.R.T.H., P.O.Box 2208, 710 03 Heraklion, Greece}}

\date{\ }

\maketitle
\begin{abstract}
We study in a systematic way all static solutions
of the Goldstone model in 1+1 dimension with a
periodicity condition imposed on the spatial coordinate.
The solutions are presented in terms of the standard 
trigonometric functions
and of Jacobi elliptic functions. Their stability analysis
is carried out, and the complete list 
of classically stable quasi-topological solitons is given.
\end{abstract}
\medskip
\medskip
\newpage

\section{Introduction}
The Goldstone model, the O(2) invariant Higgs model with Mexican hat
potential, is well known as
a theoretical laboratory for the study of spontaneous 
symmetry breaking in relativistic field theory.
Recently, it was pointed out that the structure of the classical
solutions of its 1+1 dimensional version on a spatial circle 
is strongly reminiscent of that of the localized solutions of 
the two-Higgs-doublet extension of the 
standard model (2HSM) of electroweak interactions \cite{bachas}. 
Several branches of its solutions and their analogs in the 2HSM
were explicitly constructed. 
Furthermore, the Goldstone model on $S^1$ is the simplest
paradigm of field theories \cite{bachas}, \cite{ba1}, \cite{bt} 
whose topological properties are too
trivial to lead to absolutely stable solitons of any kind, and 
which nevertheless support the existence of classically stable
quasi-topological solutions for some range of their parameters,
here the Higgs masses or equivalently the radius $L$ of spatial $S^1$.
The purpose of this little note is to present the complete list of 
static classical solutions of
the Goldstone model on $S^1$, the corresponding bifurcation tree and
their stability properties. 
This analysis, apart from its own mathematical interest, may provide
insight useful in our search for stable solitons in the 2HSM or in
the special case of the minimal supersymmetric standard model.

\section{The classical solutions}
In this section we present all static classical 
solutions of the Goldstone model on spatial $S^1$. The model is
defined with two real Higgs
fields $\phi_1$ and $\phi_2$, whose dynamics is described by the action
\be
\label{lag2}
{\cal L} = {1\over 2} (\partial_{\mu}\phi_1)^2+{1\over 2} (\partial_{\mu}
\phi_2)^2-V(\phi_1,\phi_2) \ \ , \ \ \mu=0,1
\ee
\be
V(\phi_1,\phi_2) ={1\over 4} (\phi^2_1+\phi^2_2-1)^2
\ee
and with periodic boundary condition on the space coordinate
$x \in [0,2\pi L]$. 
The energy functional for static configurations is given by
\be
E = \int_0^{2\pi L} dx \; \Bigl[ 
{1\over 2} ({d \phi_1 \over dx})^2 + {1\over 2} ({d \phi_2 \over dx})^2
+ V(\phi_1,\phi_2) \Bigr] \ \ .
\ee

As usual, the static solutions of this model may be thought of as
classical periodic motions of period 2$\pi$L of a
particle in two dimensions under the influence of the 
inverted potential $-V(\phi_1, \phi_2)$.
Apart from the vacuum solutions, which have $E_{vac} = 0$, 
there is a trivial solution 
\be 
\label{mex0}
\phi_1 = \phi_2 = 0 \ \ , \ \  E_0 = {L \pi \over 2}
\ee
which exists for
all values of $L$. The corresponding small oscillation 
eigenmodes, labelled by $j$, have
\be
\label{mode0}
\tilde \omega^2(j) = {1\over  L^2} (j^2 - L^2) \ \ , \ \ j=0,1,2,\dots \ .
\ee
They are four times degenerate, apart from the $j=0$ one 
which is doubly degenerate.
This solution is always unstable since $\tilde \omega^2(0) =-1/2$.

Many additional solutions, some of which were discussed in \cite{bachas}, 
\cite{ba1} 
bifurcate from the solution $\phi_1=\phi_2=0$ at critical 
values of $L$.
The generic solution of the model
(\ref{lag2}) can be expressed in terms of Jacobi elliptic 
functions; it has the form
\be
\label{mexgen}
\phi_1 = \sqrt R \cos \Omega \ \ \ , \ \ \ 
\phi_2 = \sqrt R \sin \Omega
\ee
\bea
    &R(x)&= a_1 + a_2 \,{\rm sn^2} \Bigl(\sqrt 2 \Lambda( x - x_0), k \Bigr)
\nonumber \\
    &\Omega(x)& = C \int_{\xi}^x {1\over R(y)} \ dy \ \ 
\label{mexgen2}
\eea
where $a_1, a_2, \Lambda, \xi, k, C, x_0$ are constants, 
while sn denotes the Jacobi elliptic function;
${\rm sn}(z,k)$, ${\rm sn}(z,k)^2$ are periodic functions on the 
real axis with periods $4K(k)$ and $2K(k)$, respectively.
$K(k)$ is the  complete elliptic integral of the first kind. 
Inserting (\ref{mexgen}) and (\ref{mexgen2}) into the equations 
corresponding to (\ref{lag2}) leads to 
several conditions on the parameters and correspondingly
to the following three types of non-trivial periodic 
solutions:

\subsection{Type-I solution}
For $a_2 = 0$ the function $R(x)$ is a constant
and the solution reduces to the form
\be
\label{mex1}
 \phi_1 = \sqrt{1-{N^2\over {L^2}}} \cos 
\left({N x\over L}+ \theta \right)
\ \ \ , \ \ \ 
\phi_2 = \sqrt{1-{N^2\over {L^2}}} \sin \left({Nx\over L}+ \theta \right)
\ee
where $N$ is an integer and $\theta$ an arbitrary constant;
it was first obtained in \cite{ba1}.
This solution  bifurcates from (\ref{mex0})
at $L=N$ i.e. when one of the $\tilde \omega$ of Eq. (\ref{mode0}) 
crosses zero. 
The Higgs field winds $N$ times around the top of the Mexican hat.
Its energy is given by
\be
     E_I(L,N) = {\pi N^2 \over 2 L} (2 - {N^2 \over L^2})
\ee

\subsection{Type-II solution}
For $C=0$ the function $\Omega(x)$ vanishes,
the parameter $a_1$ vanishes as well and the solution
takes the form
\be 
\label{mex2}
\phi_1 = 2 \ k \ \Lambda \ {\rm sn}(\sqrt 2 \Lambda x, k) \ \ \ , \ \ \ 
\phi_2 = 0 \ \ \ , \ \ \ \Lambda^2 = {1 \over 2(1+k^2)}
\ee
The Higgs field oscillates in the $\phi_2 = 0$ plane about the 
origin $\phi_1 = 0 = \phi_2$.
Evidently, this solution is  
the Manton-Samols sphaleron \cite{ms} embedded into the Goldstone model.
The argument $k$ of the Jacobi elliptic function sn has to be chosen
such that the periodicity condition is fullfilled, i.e. 
\be 
      L = {2 K(k) m \over \pi}  \sqrt{1+k^2} \ \ \ , \ \ 
\ee
for some integer $m$. When  $k \rightarrow 0$ (i.e. $L \rightarrow m$)
the solution (\ref{mex2}) approaches (\ref{mex0}).
Note that $x$ can be translated by a constant and that
the field $\phi_1 + i \phi_2$ can be rotated by a constant phase.
The energy $E_{II}$ is identical to the one of the
one-Higgs model \cite{ms}. The relevant integral reads
\be
\label{eneII}
     E_{II}(L,m) = {8 m \over \sqrt{2} \Lambda (1+k^2)^2}
         \int_0^{K(k)} dy
    \ \Bigl[ \Bigl(k^2 {\rm sn}^2(y,k) - {1+k^2 \over 2} \Bigr)^2
             + {2k^2-1-k^4 \over 8} \Bigr]
\ee
and can be evaluated in terms of the elliptic functions K(k), E(k)
by means of the integrals
\bea
\label{iden}
   &\int_0^{K(k)} dy \ {\rm sn}^2(y,k) &={K-E \over k^2} \nonumber \\
   &\int_0^{K(k)} dy \ {\rm sn}^4(y,k) &={(2+k^2)K - 2(1+k^2)E \over 3 k^4}   
\eea

In particular
\bea
&{\rm for \ \ }k=0 \ \ &E_{II}(L=m,m) = {m \pi \over 2}
 \nonumber       \\
&{\rm for \ \ }k=1 \ \ &E_{II}(L=\infty,m) = {8 m \over 3 \sqrt{2}}   
\eea

\subsection{Type-III solution}
When all the parameters entering in (\ref{mexgen}), (\ref{mexgen2}) 
are non zero the solutions are more involved. They were mentioned
without details in \cite{bachas} and we construct them here, as explicitely 
as possible. The equations imply the conditions 
\be
\label{mex3}
    a_1= {2 \over 3}(1 - 2 \Lambda^2 (1+k^2)) \ \ , \ \ 
    a_2=  4 k^2 \Lambda^2  
\ee
\be    
C^2 = {4\over 27} 
\Bigl(1+(4k^2-2)\Lambda^2 \Bigr) 
\Bigl(1+(4-2k^2)\Lambda^2 \Bigr)
\Bigl(1-(2k^2+2)\Lambda^2 \Bigr)
\ee
leaving a family of solutions depending on the four parameters~:  
$k,\Lambda,\xi,x_0$.
The condition that $R$ and $C^2$ should be positive implies
\be
\label{limite}
          \Lambda^2 \leq {1 \over 2(1+k^2)}
\ee
When the equality holds one is led to the type-II solutions
discussed above.

In order for $\Omega$ and $R$ to be periodic on $[0,2\pi L]$
the following conditions must be satisfied
\be
\label{periodo}
     C \int _0^{2\pi L} {1 \over R(y)} dy = 2 \pi n 
\ee
\be  
\label{periodr}
           L = {m K(k) \over \sqrt{2} \pi \Lambda}
\ee
for some positive integers $m,n$.
These conditions
fix $\Lambda$ and $k$ as functions of $L,m,n$. 
Thus, for a given L, the generic solution depends
on the parameters $\xi$, $x_0$, $m$ and $n$. The first two 
correspond to the arbitrary global
phase and position of the configuration. The integer 
parameters $m$ and $n$
determine respectively the number of oscillations of the modulus of 
the Higgs field and the number of the Higgs field windings 
around the origin $\phi_1 = 0 =\phi_2$ in a period $2\pi L$.   

Solving (\ref{periodo}), (\ref{periodr})
in the case $k=0$ (with $K(0)=\pi/2$) we find easily 
the critical values of $L$ where the type-III solutions 
start to exist~:
\be
\label{lcrit}    
  \Lambda^2 = {m^2 \over 4(6n^2-m^2)} \ \ \Rightarrow \ \   
   L^2 = {1 \over 2} (6 n^2 - m^2)
\ee
The expression for $\Lambda^2$ above combined with (17) 
lead to the condition $2n > m$ on the integers $m$ and $n$.
The positivity of $L^2$ in (20) then follows automatically.
This result also demonstrates that for $n$ fixed 
there are $2n-1$ possible values of $m$.

Due to the absence of a closed form for the integral
\be
\label{inte}
         \int_0^{K(k)} {1 \over c+ {\rm sn}^2(y,k)} dy
\ee
for generic values of $k$, the condition (\ref{periodo})
is impossible to handle algebraically. 
Even so, one can make some progress with
the analysis by means of a $k^2$ expansion. 

For any $n$ and $m \neq 2n$ the coefficient
of sn$^2$ in the integral giving $\Omega(x)$ in (\ref{mexgen2})
is proportional to $k^2$ and one may easily expand the solution
in powers of $k^2$. We find
\bea 
\label{expa}
  \Lambda^2 &=& {m^2 \over 4(6n^2-m^2)} \Bigl(1 + {k^2 \over 2}\Bigr) 
  + {\cal O}(k^4)
\nonumber \\
  L^2 &=& {6n^2 - m^2 \over 2} + {\cal O}(k^4) \nonumber \\
C^2 &=& {2 n^2 (m^2-4n^2)^2 \over (6n^2-m^2)^3} + {\cal O}(k^4)
\eea
Inspection of the limit $k=0$
shows that the $2n-1$ solutions of type-III  bifurcates from the type-I 
solution with $N=n$ at $L^2 = (6n^2-m^2)/2$, for $m=1,2,\dots 2n-1$.

The energy can also be expanded in powers of $k^2$ leading to
\be
     E_{III}(k,m,n) = {\pi \sqrt{2}n^2 
                     \over (6n^2-m^2)^{3/2}}
\Bigl(5n^2-m^2 - k^4{3 m^2(12n^4-m^2n^2-m^4)\over 64(6n^2-m^2)(4n^2-m^2)} 
+ {\cal O}(k^6) \Bigr)
\ee
The dependence on $L$ is recovered  by (\ref{expa}). 
Correspondingly, the $k^2$-expansion of the energy of the Type-I 
solution about the point $L^2 = (6n^2-m^2)/2$ leads to
\be
     E_{I}(k,N=n) = {\pi \sqrt{2} n^2 
                     \over (6n^2-m^2)^{3/2}}
\Bigl(5n^2-m^2 - k^4{3 m^2(4n^2+m^2)(3n^2-m^2)\over 64(6n^2-m^2)(4n^2-m^2)}
 + {\cal O}(k^6) \Bigr)
\ee
The case $m=n=1$ corresponds to the
branch labelled $\tilde W_1$ in \cite{bachas}.
In this case the energies $E_I$ and $E_{III}$ deviate only
from the $k^8$ term on~:
 \be
    E_I(k,1) = \pi {8\over 25} \sqrt{{5\over 2}}
   \Bigl[ 1 -{1 \over 128}(k^4 + k^6) - {2105\over 294912}k^8 
   + {\cal O}(k^{10}) \Bigr]
\ee
\be
 E_{III}(k,1,1) = \pi {8\over 25} \sqrt{{5\over 2}}
\Bigl[ 1 -{1 \over 128}(k^4 + k^6) - {2045\over 294912}k^8 
+ {\cal O}(k^{10}) \Bigr]  \ 
\ee
Thus $E_I(k,1)$ is lower than $E_{III}(k,1,1)$ but only very slightly, as
pointed out in \cite{bachas} on the basis of a numerical study of
these solutions.

We have studied Eqs. (\ref{periodo}), (\ref{periodr}) 
numerically, solving for $\Lambda^2$ as a function of $k^2$
for different values of $n/m$. For fixed values of $k, n/m$
we find a single solution $\Lambda^2(k,n/m)$ obeying the 
following property 
\be
\label{lambda2}
       \Lambda^2(k=0,n/m) = {m^2 \over 4(6n^2-m^2)} \ \ \ , \ \ \ 
       \Lambda^2(k=1,n/m) = {1 \over 4}
\ee
This is illustrated in Fig.[1] for $n/m=1$ and $n/m =4/7$ by the solid lines,
the dashed line representing the limit (\ref{limite}). 
The second limit (\ref{lambda2}) and the form of the
solutions suggest that in the limit 
$k \rightarrow 1$, which corresponds to $L\rightarrow \infty$,
solution III aproaches solution II. 
To test this statement, it is interesting to compare 
their energies. The energy of the type-III solution is given by the integral
\be
     E_{III}(k,m,n) = {m \over \sqrt{2} \Lambda}
     \int_0^{K(k)} dy
   \Bigl[ \Bigl(a_1 - 1 + a_2 \,{\rm sn}^2(y,k)\Bigr)^2
         + {1 \over 6} \Bigl(1 + 16 \Lambda^4(k^2-1-k^4) \Bigr) \Bigr]
\ee
For $k=1$ one obtains  
\be 
         E_{III}(1,m,n) = {4 m \over 3 \sqrt{2}} = {1 \over 2} E_{II}(1,m)
\ee 
This verifies our expectation that solution III approaches solution
II in the limit $k\rightarrow 1$.
The occurence of the factor $1/2$ is due to the fact that the solution
III depends on sn$^2$ which has a period $2 K(k)$ while the 
solution II depends on sn whose period is $4 K(k)$.

The energies of the solutions of a few low lying branches
are plotted in Fig. [2] as functions of $L$. For $L > \sqrt{5/2}$
all four types of solutions coexist and satisfy 
\be
      E_I(L,N=1) < E_{III}(L,m=1,n=1) < E_{II}(L,m=1) < E_0(L)
\ee 
The circle shows the 
bifurcation value $L^2 = 5/2$   
of the $n=m=1$ type-III solution from the $N=1$ type-I
solution. The stars indicate the three bifurcating values
($L^2 = 15/2, 10, 23/2$) of the $n=2$, $m=1,2,3$ type-III
solution from the $N=2$ type-I solution
(energies of the $n=2$ type-III solutions are not plotted).
The numbers in parentheses in the figure represent
the number of negative modes and the number of zero modes 
respectively of the corresponding solution.
They follow from the stability analysis which is the content
of the next section.

\section{Stability}

\subsection{Type-I solutions}
We start with the stability analysis of the type-I solutions (\ref{mex1}).
This set contains the lowest energy non-trivial solution, the 
branch $W_1$ of \cite{bachas}, which was shown in \cite{ba1} to be 
for $L > \sqrt{5/2}$ classically stable and 
therefore to be a soliton of the model.
Do there exist more solitons in this class of solutions?

To analyse the stability of these solutions 
we will adopt the point of view of hidden algebra and
of differential operators preserving finite dimensional 
spaces of polynomials  \cite{bgkk}, which in the case
at hand is a rather straightforward application of Fourier analysis.

Perturbing as usual the fields $\phi_a, \,a=1,2$ 
around the classical solution 
(\ref{mex1}), denoted here by $\phi_a^{\rm cl}$,
\be
        \phi_a(x) = \phi_a^{\rm cl}(x) + \eta_a(x) \exp (- i \omega t)
\ee
leads to the following equation for the normal modes~:
\be
A(N,L)
\left(\begin{array}{c}
\eta_1\\
\eta_2
\end{array}\right)
= \omega^2
\left(\begin{array}{c}
\eta_1\\
\eta_2
\end{array}\right)
\ee
\be
\label{sch}
A(N,L) \equiv - {d^2 \over dx^2}
+ 2\Bigl(1-{N^2\over {L^2}} \Bigr)
\left(\begin{array}{cc}
c^2 &sc\\
sc &s^2
\end{array}\right)
- {N^2\over {L^2}}
\ee
where $\omega^2$ is the eigenvalue and, to simplify notation, 
we posed $c=\cos(N x / L)$ and $s = \sin(N x / L)$.

The complete list of eigenvalues of the operator $A(N,L)$ for $N \geq 1$
can be obtained by classifying its invariant subspaces.
This can be done by using the Fourier decomposition. After some algebra,
one can check that for an integer $n \geq N$
the following finite dimensional vector
spaces are preserved by  $A(N,L)$ 
\bea
&V_n =&{\rm Span} \lbrace
        \alpha_p \cos{px \over L} \ + \ \beta_p \sin{px \over L},
        \ \ p-n = 0\ ({\rm mod}\ 2N), \ \ \vert p \vert \leq n 
              \rbrace \ \ , \nonumber \\
&\tilde V_n =&{\rm Span} \lbrace
        \alpha_p \sin{px \over L} \ - \  \beta_p \cos{px \over L},
        \ \ p-n = 0\ ({\rm mod}\ 2N), \ \ \vert p \vert \leq n 
              \rbrace
\eea
provided
\be
     \alpha_p = \beta_p \ \ \ {\rm if} \ \ \ \ p-n+2N > 0
\ee
while the other constants $\alpha_p, \beta_p$ are arbitrary.
The operator $A(N,L)$ can then be diagonalized
on each of the finite dimensional vector space above,
leading to a set of algebraic equations. 
One finds  finally  that
the eigenvalues of $A(N,L)$ on $V_n$ read
\be
\label{modeI}
   \omega^2(N,k,\pm 1) = { 1\over L^2}
\Bigl( (L^2-N^2) + k^2 \pm \sqrt{ (L^2-N^2)^2 + 4 N^2 k^2}
\Bigl)
\ee
for $k = 0, 1, 2, \dots$ and similarly on $\tilde V_n$ 
 for $k = 1, 2, 3, \dots$.
The spectrum  given by  (\ref{modeI}) fits exactly
with (\ref{mode0}) in the limit $L=N$ for all $N$
i.e. when $L$ approaches the points of bifurcation 
of the type-I solutions from (\ref{mex0}).

We can now discuss the stability of the solutions of type-I.
First remark that 
$\omega^2(N,0,-1)=0$, it corresponds to the zero mode related to
the invariance of the equations under translations of $x$.
For $N$ fixed and $L$ slightly greater than $N$, 
the solution (\ref{mex1}) possess $4N-2$ negative modes
corresponding to $\epsilon = -1$ and
$k=1, 2, 3, \dots, 2N-1$ in (\ref{modeI}), remembering that
they are twice degenerate.
When $L$ increases, more and more of these eigenmodes become 
positive, crossing zero at the values
\be
     L^2 = {1 \over 2} (6N^2 - m^2) \ \ \ , \ \ \ m = 1,2, \dots 2N-1
 \ \ ,
\ee
i.e. (cf. (\ref{lcrit})) at those values of $L$ where the 
solutions of type-III
with $n=N$ bifurcate from the solution of type-I. For
\be
      L^2 \geq L^2_{cr}(N) \equiv {1 \over 2} (6N^2 - 1)
\ee 
all the modes are positive and (\ref{mex1}) 
are classically stable solitons.
These results are illustrated on Fig.[2] for $N=1$ and $N=2$.
The numbers in parenthesis represent the number of negative
and of zero modes of the corresponding branch.

\subsection{Alternative approach for Type-I}
The stability analysis could also be performed by
considering the polar decomposition of the Higgs fields \cite{ba1}
\be
\label{polar}
\phi_1 = F\cos \Theta\quad , \quad \phi_2 = F\sin \Theta
\ee
The quadratic operator $Q$ associated with this paramerization
of the fields admits the following invariant subspaces
\begin{eqnarray}
\label{basis}
U_n &=& {\rm{span}} \lbrace {1\over{\sqrt{\pi L}}} (\cos {nx\over L},0),
{1\over{\sqrt{\pi L}}} (0, \sin {nx\over L})\rbrace\\
\tilde U_n &=& {\rm{span}} \lbrace {1\over{\sqrt{\pi L}}} (\sin
{nx\over L}, 0), {1\over{\sqrt{\pi L}}} (0, \cos {nx\over L} x)\rbrace
\end{eqnarray}
which, by Fourier theory, cover the whole relevant
Hilbert space of periodic functions.

In the  {\it normalized} basis (\ref{basis}), the eigenvalues
of $Q$ on  the subspace $U_n$ read
\begin{eqnarray}
\label{eigenval}
\omega^2(N,n,\pm 1) &=& {1\over{2L^2}} \lbrace n^2 L^2 
+\Delta^2(n^2+2L^2)\nonumber\\
&\pm&
\sqrt{\Delta^4(n^4-20n^2L^2+4L^4)+2n^2L^2\Delta^2(10L^2-n^2)+n^4L^4}\rbrace
\end{eqnarray}
where we defined $\Delta^2 = L^2-N^2$. The eigenvalues corresponding 
to $\tilde U_n$
are identical to (\ref{eigenval}). 

These values fail to approach the
$\tilde \omega^2$ of Eq.(\ref{mode0}) in the limit $L \rightarrow N$.
This is due to the fact that the  parametrization (\ref{polar}) 
is singular about
the solution $\phi_1=\phi_2=0$. 
Remarkably though, the values (\ref{modeI}) and (\ref{eigenval})
have zero crossing at exactly the same values of $L^2$
and the conclusions about the stability of the solutions 
are identical in the two approaches.

\subsection{Type-II and III solutions}
The stability analysis about the solution of type-II can immediately be
carried out. The equations for the fluctuations about
the solution (\ref{mex2}) decouple to take the form of the Lam\'{e}
equations:
\bea
\label{lame6}
\lbrace  &- {d^2 \over dy^2}
     + 6  k^2 {\rm sn}^2(y,k) 
\rbrace \eta_1 &= \Omega_1^2 \eta_1  \\
\label{lame2}
\lbrace  &- {d^2 \over dy^2}
     + 2  k^2 {\rm sn}^2(y,k)
\rbrace \eta_2 &= \Omega_2^2 \eta_2
\eea
where we posed 
\be y = \sqrt{1+k^2} x \ \ , \ \ 
\Omega_a^2 \equiv (\omega_a^2+1)(k^2+1) \ \ , \ \ a=1,2
\label{lamedef}
\ee
$\omega_a^2$ being the effective eigenvalue of the relevant operator. 
Equations (\ref{lame6}) and (\ref{lame2}) admit five and three algebraic
modes respectively, with corresponding eigenvalues
\be
\Omega_1^2 \ \  : \ \ 4 + k^2 , 1 + 4 k^2 , 1+ k^2 , 
  2(1+k^2) \pm 2\sqrt{1-k^2+k^4}
\label{lame5}
\ee
\be
\label{lame3}
\Omega_2^2 \ \ : \ \ 1 + k^2 , 1 , k^2
\ee
The corresponding values of $\omega^2$ follow immediately from
(\ref{lamedef}); they have signature $(+,+,0,+,-)$ and $(0,-,-)$
respectively for (\ref{lame5}) and (\ref{lame3}).
Each of the
equations (\ref{lame6}),(\ref{lame2}) therefore leads to a zero mode of the 
solution (\ref{mex2}), their origin was discussed in Sect.2.
It is a property of the Lam\'{e} equation that the solutions determined
algebraically correspond to the solutions of lowest eigenvalues.
The remaining part of the spectrum therefore consists of positive
eigenmodes.
The spectrum of Eq.(\ref{lame6}) was studied  perturbatively 
in \cite{ms}, while the relation between the Lam\'{e} equation and the 
Manton-Samols sphalerons was first pointed out in \cite{muller}.
The presence of negative modes in the small oscillation spectrum
of all type-II solutions means that no classically stable soliton
exists among them.

The stability equation associated with the type-III
solutions seems to be more difficult, mainly due to
the lack of analytical expression for $\Omega(x)$ in Eq. (\ref{mexgen}). 
We have not
obtained convincing algebraic expressions for their
normal modes but a few of their properties can be pointed out.
First, these solutions should possess a double zero mode due to 
the invariance of the solution (\ref{mexgen}), (\ref{mex3})
under translations of $x$ and internal O(2) rotation. 
Likely the eigenvalues relative to these solutions connect
to the values (\ref{modeI}) in the limit 
$L^2 \rightarrow  (6n^2-m^2)/2$, $N=n$.
We conjecture that the solutions of type-III always possess  
at least one negative mode. In the case $n=m=1$,
the negative mode should be unique and meet the two zero modes
at the point $L= \sqrt{5/2}$, $\omega^2 = 0$, indicated by
the star of Fig. [3]. 

\section{Conclusion}
The model (\ref{lag2}) on $S^1$ admits a rich set of 
solutions which bifurcate from each other at
critical values of the spatial radius $L$.
A detailed analytical study of these solutions was presented.
Their stability analysis was carried out, all
classically stable solitons were identified, together with the
range of the parameter $L$ for which they are stable.
The bifurcation pattern contains as a subset the 
branches corresponding to the solutions found in \cite{ms}, and it 
is found to be richer than that of the gauged Higgs model \cite{bgkk},
the gauged version of (\ref{lag2}). 

The spectrum of modes of small fluctuations around a classical
solution is shown to depend on the parametrization employed for
the complex scalar. In particular, we explicitly demonstrated
this fact for the normal modes around the type-I solutions.
As one varies the parameter $L$ to approach the bifurcation point
$L=N$ along the type-I solution with winding $N$, the correct
spectrum is the one which coincides with that of the main branch
(\ref{mex0}). 

In this work we concentrated on static solutions. The study of 
time dependent ones, to search for the analogs of the 
breather solutions of the sine-Gordon equation,
is particularly interesting. They may well be the prototypes 
of stable "breathing membranes" in realistic particle physics
models.


\newpage
\centerline{Figure Captions}
\begin{itemize}
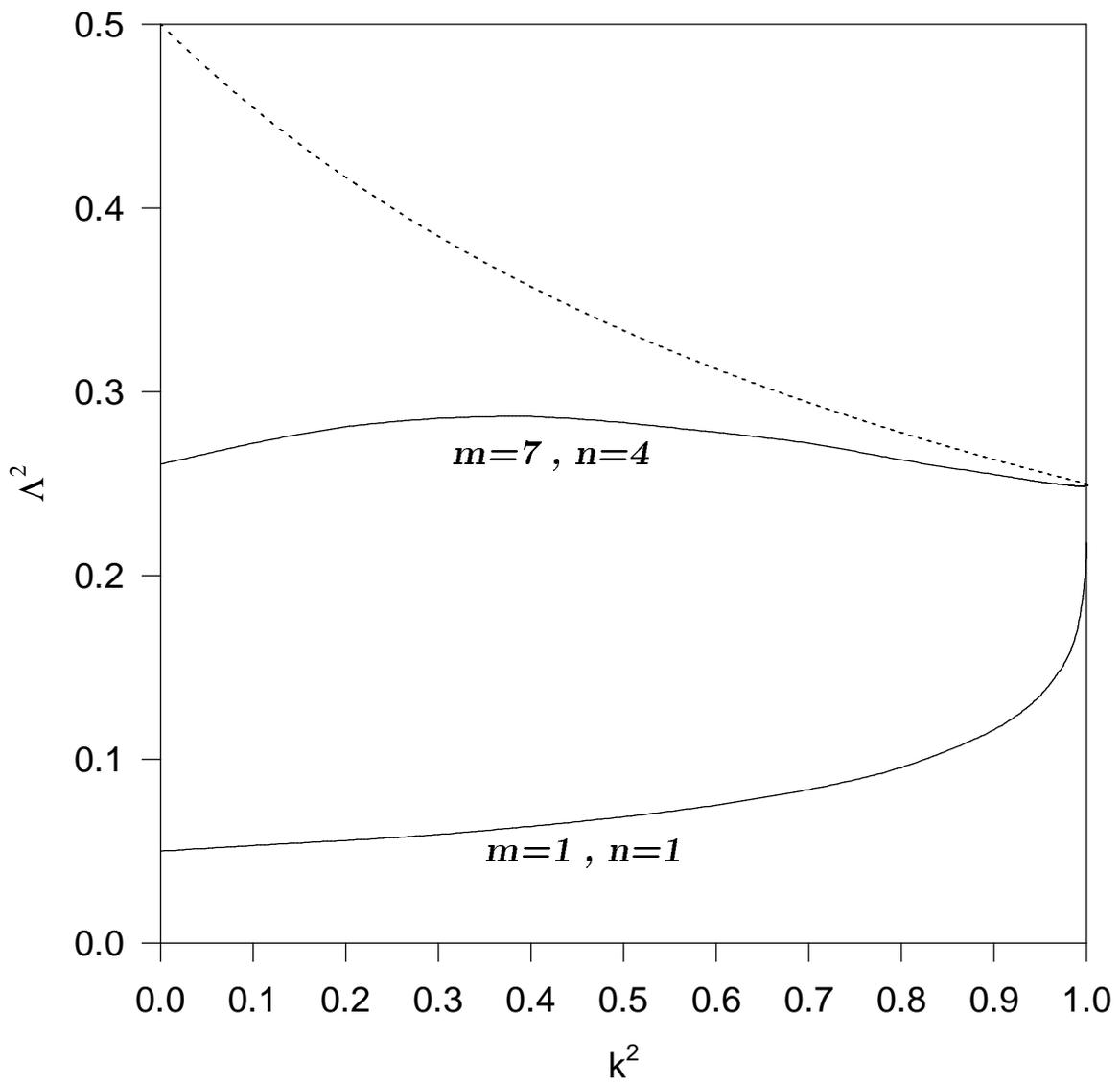
\item { } {\bf Figure 1.}
The solutions of eqs. (\ref{periodo}), (\ref{periodr})
are plotted as functions of $k^2$ for two values of $n/m$.
The dashed line indicates the limit (\ref{limite}).
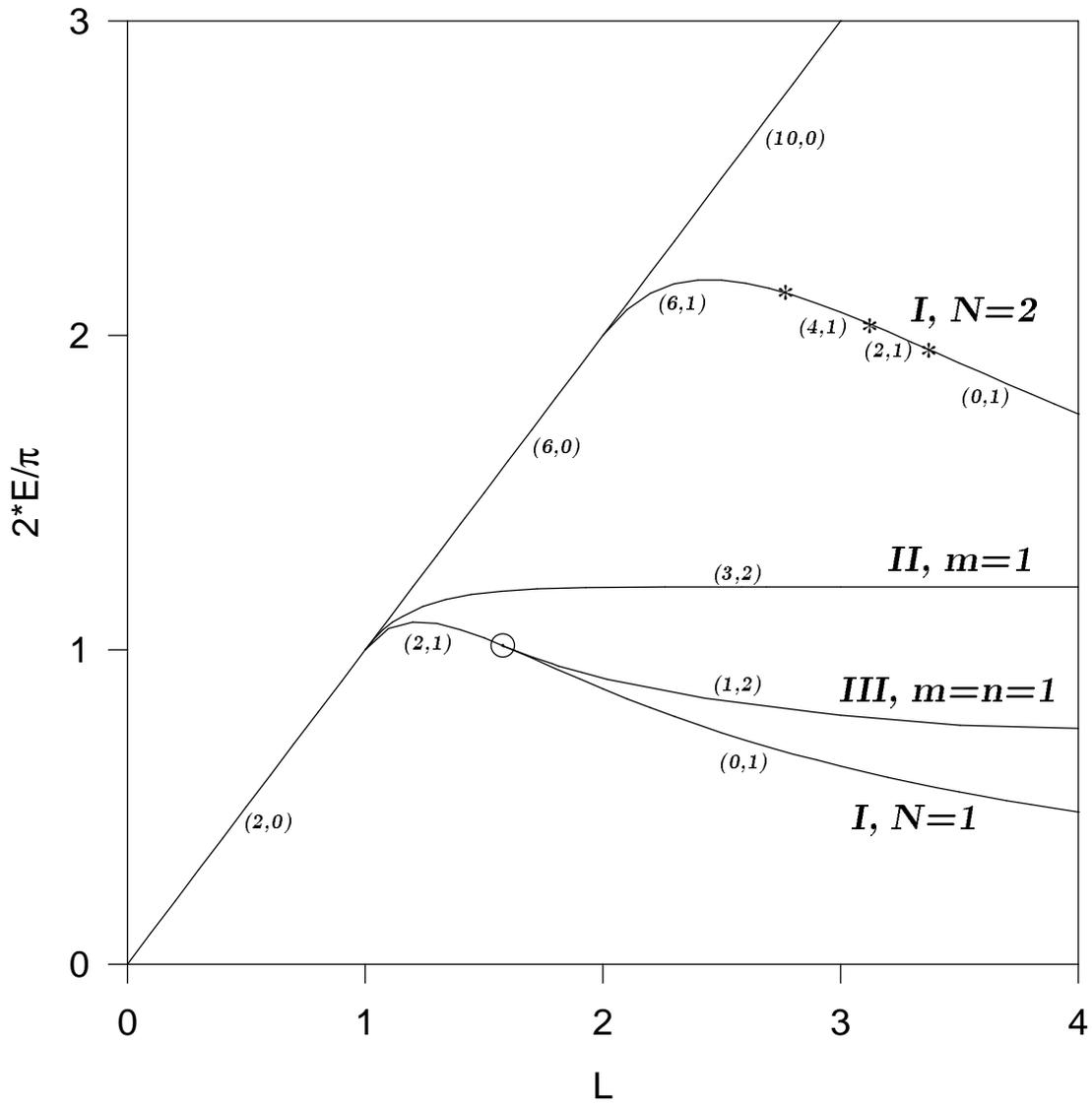
\item { } {\bf Figure 2.}
The energies of some of the solutions are plotted 
as functions of the parameter $L$.  The circle indicates the 
bifurcation point ($L^2 = 5/2$) on  the $N=1$ type-I solution. 
The stars indicate
the three bifurcation points ($L^2 = 15/2,10,23/2$)
on the $N=2$ type-I solution. The two numbers in parentheses
refer to the number of negative modes and of zero modes of
the corresponding solution.
\item { } {\bf Figure 3.}
The  values (\ref{mode0}) are
plotted as functions of $L$ for $j=0,1,2$ (solid lines),
together with the values 
$\omega^2(1,1,-1)$,
$\omega^2(1,0,-1)$,
$\omega^2(1,0,1)$,
$\omega^2(1,2,-1)$,\\
$\omega^2(1,1,1)$,
$\omega^2(1,3,-1)$
of (\ref{modeI})  (dotted lines).
The numbers indicate the multiplicity of the eigenvalues.
\end{itemize}
\end{document}